# Ultrafast unbalanced electron distributions in quasicrystalline 30° twisted bilayer graphene

T. Suzuki[1,*], T. Iimori[1], S. J. Ahn[2], Y. Zhao[1], M. Watanabe[1], J. Xu[1], M. Fujisawa[1], T. Kanai[1], N. Ishii[1], J. Itatani[1], K. Suwa[3], H. Fukidome[3], S. Tanaka[4], J. R. Ahn[2,5], K. Okazaki[1], S. Shin[1], F. Komori[1,*], and I. Matsuda[1,*]

[1]*Institute for Solid State Physics, University of Tokyo, Kashiwa, Chiba 277-8581, Japan*

[2]*Department of Physics and SAINT, Sungkyunkwan University, Suwon, Gyeonggi-do 16419, Republic of Korea*

[3]*Research Institute of Electronical Communication, Tohoku University, Sendai, Miyagi 980-8577, Japan*

[4]*Department of Applied Quantum Physics and Nuclear Engineering, Kyushu University, Fukuoka, Fukuoka, 819-0395, Japan*

[5]*Samsung-SKKU Graphene Centre, Sungkyunkwan University, Suwon, Gyeonggi-do 440-746, Republic of Korea*

[*]To whom correspondence should be addressed. E-mail: takeshi.suzuki@issp.u-tokyo.ac.jp; komori@issp.u-tokyo.ac.jp; imatsuda@issp.u-tokyo.ac.jp.

**Layers of twisted bilayer graphene exhibit varieties of exotic quantum phenomena[1-5]. Today, the twist angle Θ has become an important degree of freedom for exploring novel states of matters, *i.e.* two-dimensional superconductivity ( Θ = 1.1°)[6, 7] and a two-dimensional quasicrystal (Θ = 30°)[8, 9]. We report herein experimental observation on the photo-induced ultrafast dynamics of Dirac fermions in the quasicrystalline 30° twisted bilayer graphene (QCTBG). We discover that hot carriers are asymmetrically distributed between the two graphene layers, followed by the opposing femtosecond relaxations, by using time- and angle-resolved photoemission spectroscopy. The key mechanism involves the differing carrier transport between layers and the transient doping from the substrate interface. The ultrafast dynamics scheme continues after the Umklapp scattering, which is induced by the incommensurate interlayer stacking of the quasi-crystallinity. The dynamics in the atomic layer opens the possibility of new applications and creates interdisciplinary links in the optoelectronics of van der Waals crystals.**

Carrier dynamics in graphene is determined by Fermions in a linearly dispersing band structure, the Dirac cones, which have successfully offered unique electronic properties and achieved many applications[10]. Non-equilibrium electronic states in a matter, generated by optical pumping, are characterized by the transient temperature. In a case of the *n*-type graphene layer (Fig.1a), typically produced on a SiC substrate in a large area, the temporal chemical potential reduces with the temperature, as shown in Fig.1b,c. This dynamical phenomenon is specific to the massless Dirac

Fermion and it is caused by keeping the charge-neutrality conditions in the linear density of states (DOS). On the other hand, an *n*-type bilayer graphene, shown in Fig.1d, on a substrate has massive Fermions with the energy gap, formed by interband interaction of Dirac states between the two layers. The DOS becomes constant and, thus, a position of the chemical potential does not depend on the temperature (Fig. 1e,f). For the QCTBG, the interlayer interaction is the minimum due to the 30° twisted angle and the massless Dirac Fermions remain in the two layers. As shown in Fig.1g, the electronic state results in the twelve-fold (dodecagonal) structure and it is composed of upper- and lower-layer Dirac cones, ULDs and LLDs, with their replicas emerged due to the Umklapp scattering by the interlayer stacking (Fig. S1). The QCTBG doubles two-dimensional number density of Dirac Fermions that follows the carrier dynamics in Fig.1 b,c. Here , we discover that the transient chemical potential is distinctive and it actually behaves opposite between ULD and LLD, as shown in Fig 1h and 1i while each Dirac Fermion follows the carrier dynamics in Fig.1 b,c . The main subject of this work is to share our observation of the novel phenomenon in dynamics in the QCTBG and to provide a possible origin of the contrasting ultrafast behaviour.

The unique electronic structure of QCTBG makes it ideal for revealing novel physical properties that may lead to new applications in graphene technology. We explore this electronic structure by focusing on the ultrafast dynamics through time- and angle-resolved photoemission spectroscopy (TARPES), which allows us to directly observe the temporal evolution of fermions in the Dirac cones (see Fig.1c). The technique has already revealed dynamical events, such as the bottleneck effect[11, 12] and the supercollision process[13, 14], which are specific to Dirac fermions in single layer graphene. Thus, we apply the experiment to the quasi-periodic graphene system and compare the results with those of a periodic graphene system. The measurement used the pump-probe approach with an infrared pulse with $hv$ = 1.55 eV serving as pump and an extreme-ultraviolet pulse with $hv$ = 21.7 eV serving as probe to cover the entire 2D Brillouin zone (Fig.S2c).

Figure 2 summarizes the time-resolved photoemission band diagrams acquired at selected delay times for the ULD and LLD bands in QCTBG. The results for periodic nontwisted bilayer graphene (NTBG) are also shown for comparisons. The Dirac cones of QCTBG are *n* type at the Dirac points and are below the equilibrium chemical potential ($\mu_{eq}$). The electronic structure of NTBG is also *n* type and there is a band gap around $\mu_{eq}$ (Fig. S3)[15, 16]. After the pump pulse (intensity ~0.7 mJ/cm$^2$), the TARPES band diagram of individual bands of bilayers evolves on the femtosecond time scale. To enhance the temporal variations, the band diagrams are shown as the difference between the spectra before and after photoexcitation, where red and blue in Figs. 2g–i represent an increase and decrease in photoemission intensity, respectively. The spectral weights for all bands decrease immediately below $\mu_{eq}$ and increase above $\mu_{eq}$ at $\Delta t$ = 0.05 ps. This reflects the excitation of electrons from the occupied bands to the unoccupied bands. The resulting spectra in Figs. 2a–c are consistent with the calculated band structure. At $\Delta t$ = 0.16 and 0.32 ps, the difference intensity decreases with delay time, which corresponds to the relaxation of photoexcited carriers.

To evaluate the occupation of Dirac cones by nonequilibrium carriers, we plot energy distribution curves of the Dirac bands by integrating over momentum space. Figures 3a and 3b show a series of energy distribution curves for the ULD and LLD bands, respectively, for several pump-probe delay times $\Delta t$. The spectra are fit by the Fermi-Dirac distribution function convoluted by a Gaussian to extract the electronic temperature (Fig. 3c) and chemical-potential shift $\Delta\mu$ (Fig. 3d) for the ULD and LLD bands. The similar time evolution for temperature in Fig. 3c indicates that photoexcited carriers in the two bands take the same relaxation pathway. In contrast, the different time evolution for $\Delta\mu$ indicates that the ULD and LLD bands have the opposite behaviour after $\Delta t = 0.2$ ps (*i.e.* the ULD undergoes a negative shift whereas the LLD undergoes a positive shift). Interestingly, $\Delta\mu$ for the NTBG band remains constant at essentially zero over the same time delay. The striking difference among the three types of the Dirac cones provides clear evidence of a carrier imbalance between the ULD and LLD bands of the QCTBG on the ultrafast time scale.

We now discuss the TARPES of the ULD and LLD replica bands. Figures 3g and 3h show the time dependence of temperature and $\Delta\mu$, respectively. For the transient temperature, the temporal profiles of the replica bands are almost identical to each other and to those of the original bands (Fig. S4), which indicates that, during relaxation, photoexcited electrons occupying Dirac cones interact via electron-phonon interactions with the common phonon bath. The nonequilibrium chemical potential $\Delta\mu$ of the LLD replica band is positive and decreases monotonically, as it does for the original LLD band. In contrast, $\Delta\mu$ for the ULD replica band is initially positive and, for $\Delta t > 0.3$ ps, becomes slightly negative and then approaches zero. The sign inversion of the ULD replica band differs significantly from the dynamics of the original ULD band, which implies that the electron distribution between the replica Dirac cones and the original Dirac cones is also unbalanced.

In general, the temperature increases and $\Delta\mu$ is negative for $n$ type graphene after photoexcitation, as reported previously[17]. Thus, the temporal evolution of the ULD (replica) band is natural. Conversely, the positive $\Delta\mu$ of the LLD (replica) band is quite striking. As previously mentioned, the chemical potential is determined by temperature and carrier density for Dirac fermions. Thus, the observed difference in chemical potentials stems from the asymmetric carrier transport between the upper layer (UL) and lower layer (LL). The spatial relationship among layers in QCTBG is schematically shown in Fig. 4a. In order to gain quantitative insight, we first extract the change of electron densities, $\Delta n_{\text{el}}$, for the UL and LL shown as markers in Fig. 4b from the experimentally determined time-dependent electron temperature and chemical potential (Figs. 3 c and 3d). Then, we perform calculations by solving rate equations for carrier transport. (The details of calculation is described in Supplementary Information). The main parts of rate equations are as follows,

$$\frac{dn_{\text{el}}^{\text{UL}}}{dt} = -\frac{n_{\text{el}}^{\text{UL}}}{\tau_{\text{UL}}} + \gamma_1\left(n_{\text{el}}^{\text{LL}} - n_{\text{el}}^{\text{UL}}\right) + G_1\exp\left(-\frac{t^2}{T_p^2}\right), \quad (1)$$

$$\frac{dn_{\text{el}}^{\text{LL}}}{dt} = -\frac{n_{\text{el}}^{\text{LL}}}{\tau_{\text{LL}}} - \gamma_1\left(n_{\text{el}}^{\text{LL}} - n_{\text{el}}^{\text{UL}}\right) - \gamma_2\left(n_{\text{el}}^{\text{LL}} - n_{\text{el}}^{\text{Sub}}\right) + G_2\exp\left(-\frac{t^2}{T_p^2}\right), \quad (2)$$

, where $n_{\text{el}}^{\text{UL}}$ and $n_{\text{el}}^{\text{LL}}$ are electron densities in the ULD and LLD of a twisted-bilayer graphene, respectively. $\tau_{\text{UL}}$ and $\tau_{\text{LL}}$ are lifetimes for Dirac fermions in a single-layer graphene for the UL and

LL, respectively, which are determined by hot carrier dynamics. $\gamma_1$ and $\gamma_2$ are rate constants of carrier transfer between the UL and LL, and the LL and substrate. $G_1$ and $G_2$ are coefficients of pump-induced net density flux to the UL and LL from substrate, respectively, where $T_p$ corresponds to the time resolution of our experiment (70 fs in the full width at half maximum). The situation is depicted in Fig. 4a. To reproduce the best agreement, we set $\gamma_1 = 1.5$ ps$^{-1}$, $\gamma_2 = 0.5$ ps$^{-1}$, $G_1 = 5 \times 10^{13}$ cm$^{-3}$ps$^{-1}$, $G_2 = 8 \times 10^{13}$ cm$^{-3}$ps$^{-1}$, and calculation results are shown as solid lines in Fig. 4b.

We find that the experimental results are reproduced when a value of $\gamma_1$ is large than $\gamma_2$, indicating that carrier transfer is much frequently induced between graphene layers than between the UL and substrate. Most importantly, the apparent $G_1$ and $G_2$ values demonstrate there is transient carrier doping from the substrate to the graphene layers. The unexpected carrier imbalance, Fig.4b, seems to be due to differing values between $G_1$ and $G_2$, which dictate the initial electron distributions in the UL and LL. It is of note that the larger $G_2$ value is consistent to the closer distance of LL from the substrate. Origins of the external flux into the graphene layers can be explained due to electrons in the Si dangling-bond (DB) state at the graphene/SiC interface[18]. A possibility of the photoexcited carriers in a SiC crystal is excluded since the pumping photon energy (hv=1.5 eV) is not high enough to overcome the SiC bulk band-gap (3.3 eV). The DB bands have been known to exist at the Fermi level that locates almost bottom of the SiC conduction band[18]. Area density of Si DBs at the interface corresponds to the order of $10^{15}$ cm$^2$, while density of states in a single-layer graphene is in only the order of $10^{12}$-$10^{13}$ cm$^2$ near the Dirac point[19]. Since a number of electrons in the Si DBs (interface states) are several orders of magnitude higher than that of the graphene density of states, it is large enough to dominate the transient chemical potential.

In electronic transport, Umklapp scattering plays an intrinsically central role in increasing the resistivity at low temperature, although other effects often hinder its clear identification[20]. Recent technological advances in graphene devices have led to superlattices with small lattice mismatch; in these structures electron-electron Umklapp scattering is clearly demonstrated[21]. By inducing strong interlayer coupling, QCTBG can be regarded as an alternative route to study Umklapp scattering[8, 9]. In this respect, studying carrier dynamics by using ultrafast optical spectroscopy can complement transport studies of TBG[22], and important evidence is provided by the strong correlation of the dynamic properties revealed herein between the original bands and the replica bands for QCTBG.

The unbalanced electron distributions in the ULD and LLD bands discovered in the present study offer a novel degree of freedom that can be advantageous in optoelectronic applications using graphene. The key mechanism here is the difference in carrier transfer between the two graphene layers, which is controllable by band engineering, as done by doping[15] or intercalation[23, 24] in NTBG. In terms of applications, the transient population inversion can be augmented by using excess electron transfer from the ULD band to the LLD band. The energy scale of the observed chemical-potential shift is ~10 meV, so one can anticipate a lasing medium operating in the terahertz range, which is of technological interest[25, 26]. In the field of excitonic physics in bilayer graphene[27], asymmetric electron and hole distributions between the upper and lower layers lead to spatially separated interlayer excitons. By twisting the two layers, the electrons and holes can be separated in momentum space and in real space, as reported for the van der Waals heterostructure[28]. This will allow us to explore new directions in valleytronics in bilayer graphene.

**Methods**

**Sample preparation.** The QCTBG was grown on the Si face of a 4H-SiC (0001) substrate by thermal decomposition of the substrate in a vacuum after growing a monatomic hexagonal boron nitride layer. The upper and lower layers are rotated by 0° and 30° with respect to the orientation of the SiC (0001) surface. Details of the growth procedure are given in ref. [8]. Nontwisted bilayer graphene was grown on the Si-face of a vicinal 6H-SiC substrate by thermal decomposition in a vacuum[29]. After transferring through air to an ultrahigh-vacuum chamber, the graphene was cleaned by annealing at 450 °C to remove surface contamination.

**Photoemission measurements.** To characterise the sample, static angle-resolved photoemission spectroscopy measurements were made by using a He discharge lamp and a hemispherical electron analyser (Omicron-Scienta R4000) with an energy resolution of ~12.5 meV. For the TARPES measurements, we used for the pump pulse an extremely stable commercial Ti:sapphire regenerative amplifier system (Spectra-Physics, Solstice Ace) with a centre wavelength of 800 nm and pulse width of ~35 fs. Second harmonic pulses generated in a 0.2-mm-thick crystal of $\beta$-BaB$_2$O$_4$ were focused into a static gas cell filled with Ar to generate higher harmonics. By using a set of SiC/Mg multilayer mirrors, we selected the seventh harmonic of the second harmonic ($h\nu$ = 21.7 eV) for the probe pulse. The temporal resolution was determined to be ~70 fs from the TARPES intensity far above the Fermi level, corresponding to the cross correlation between the pump and probe pulses.

**Analysis of experiment data.** The energy distribution curves for each band were analysed by fitting a Fermi-Dirac distribution convoluted with a Gaussian such that the density of states, Gaussian width, and background remain fixed while the electronic temperature and chemical potential are varied to obtain the best fit. Errors are estimated based on the standard deviation of the fitting parameters well before the arrival of the pump pulse.

**Data availability.** The data supporting the findings of this study are available from the corresponding author.

**Acknowledgements** We would like to acknowledge Y. Tsujikawa, M. Sakamoto, and A. Takayama for their kind support during the experiments. We would also like to thank M. Koshino, T. Ishimasa, K. Kimura, and E. Minamitani for valuable discussions and comments. This work was supported by JSPS KAKENHI (Grant No. 16H06361, 18H03874, and 18K19011) and the Q-Leap, Japan. We thank Brett Kraabel, Ph.D., from Edanz Group (www.edanzediting.com/ac) for editing a draft of this manuscript.

**Author Contributions**   T.S., T.I., Z.Y., M.W., and J.X. made the TARPES measurements. T.S. analysed the data. T.S. and I.M. performed the calculations. M.F., T.K., N.I., and J.I. maintained the HHG laser system and improved the in ULD and LLD bands. Experimental data of chemical potential shifts are shown as black solid lines for ULD (**h**) and LLD (**i**), respectively.

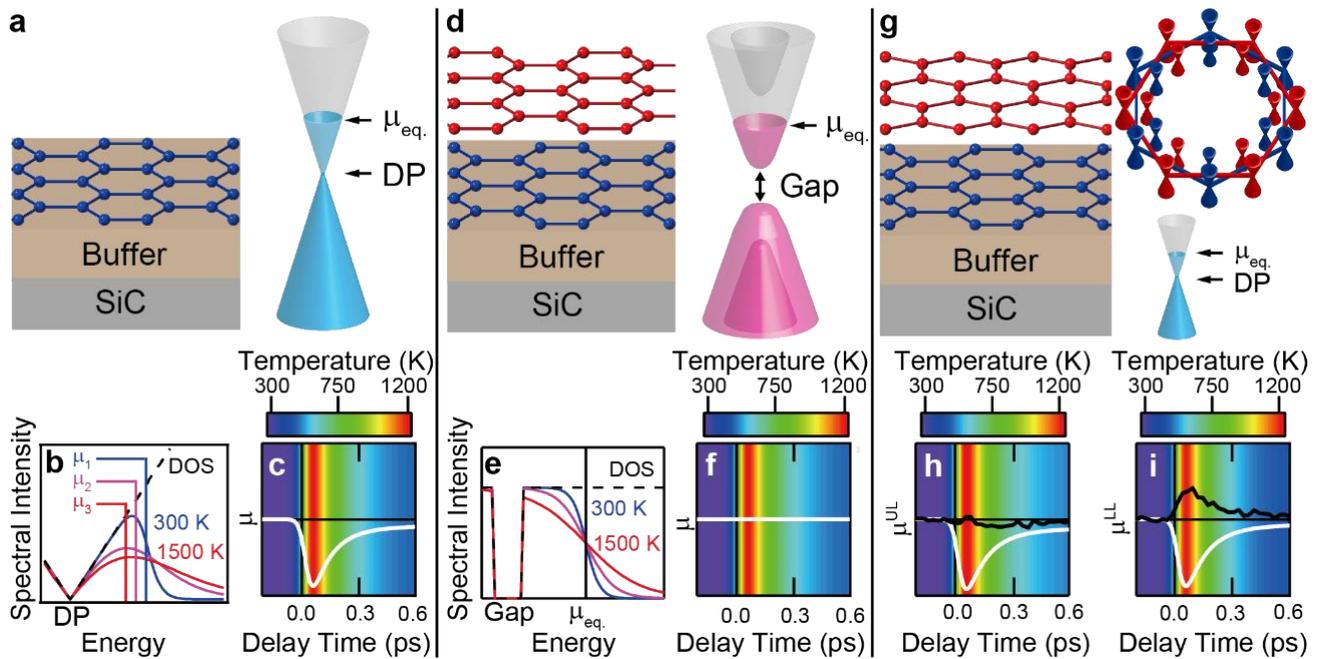

**Figure 1 | Crystal and electronic structures of n-type single-layer, bilayer, and quasicrystalline 30° twisted bilayer graphenes and hot carrier dynamics. a,** Crystal and electronic structure of single-layer graphene. Dirac point and equilibrium chemical potential are denoted as DP and $\mu_{eq.}$, respectively. **b,** Spectral intensities with different temperatures for single-layer graphene. **c,** Dynamics of temperature and chemical potential shifts for hot carriers. **d,** Crystal and electronic structure of bilayer graphene. **e,** Spectral intensities with different temperatures for bilayer graphene. **f,** Dynamics of temperature and chemical potential shifts for hot carriers. **g,** Crystal and electronic structure of quasicrystalline 30° twisted bilayer graphene (QCTBG). Outer red and blue Dirac cones represent the upper-layer Dirac (ULD) and lower-layer Dirac (LLD) bands, respectively, whereas the inner red and blue Dirac cones corresponds to the replica bands of the ULD and LLD bands. **h,i,** Dynamics of temperature and chemical potential shifts for hot carriers

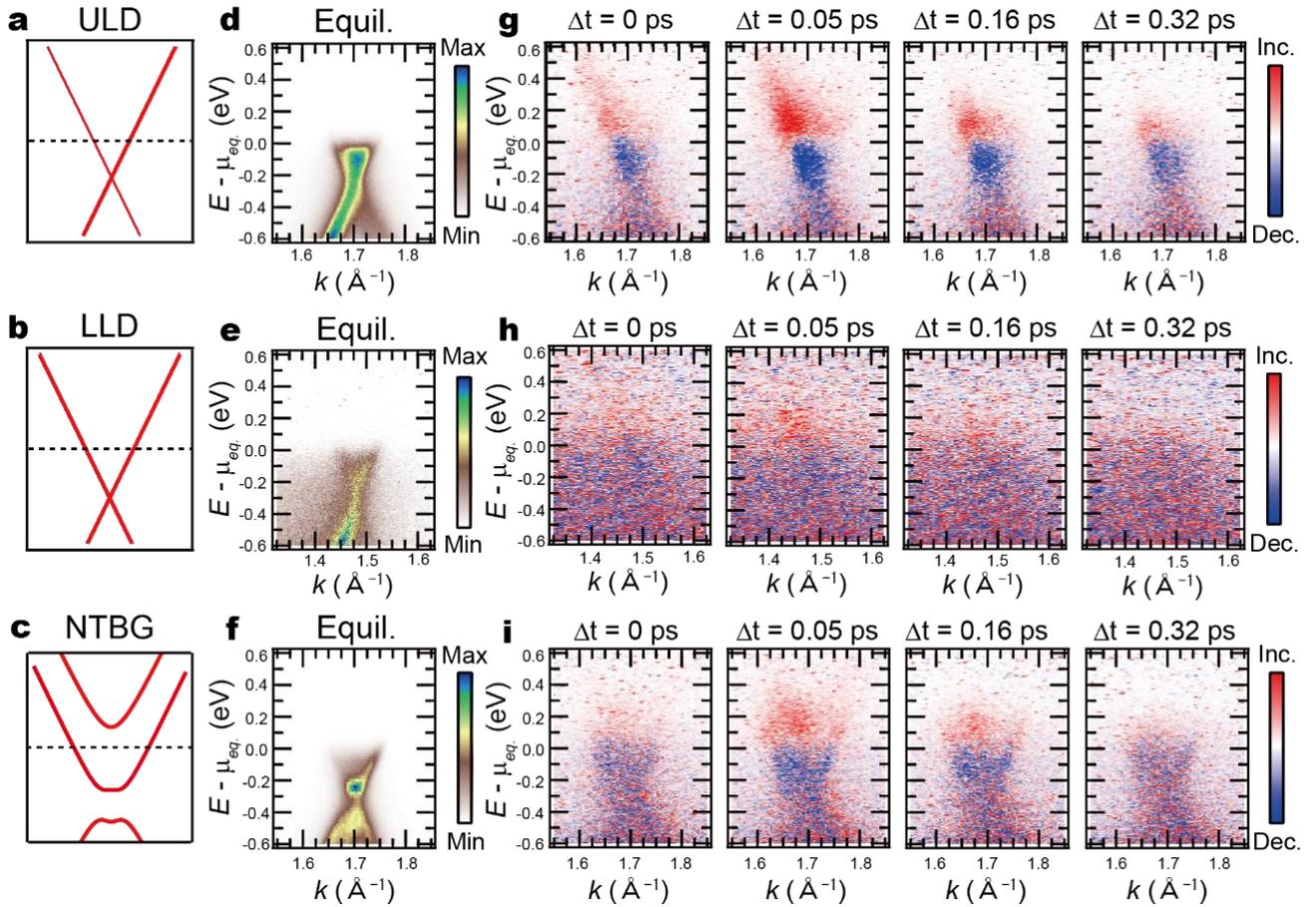

**Figure 2 | TARPES spectra for QCTBG and nontwisted bilayer graphene. a–c,** Calculated band structures for the ULD and LLD bands in QCTBG and for non-twisted bilayer graphene (NTBG) band, respectively. Dashed lines indicate the Fermi level. **d–f,** Equilibrium angle-resolved photoemission spectra for ULD, LLD, and NTBG bands. **g–i,** Difference images of time- and angle-resolved photoemission spectroscopy (TARPES) for the ULD, LLD, and NBLG bands. Red and blue points represent increasing and decreasing photoemission intensity, respectively.

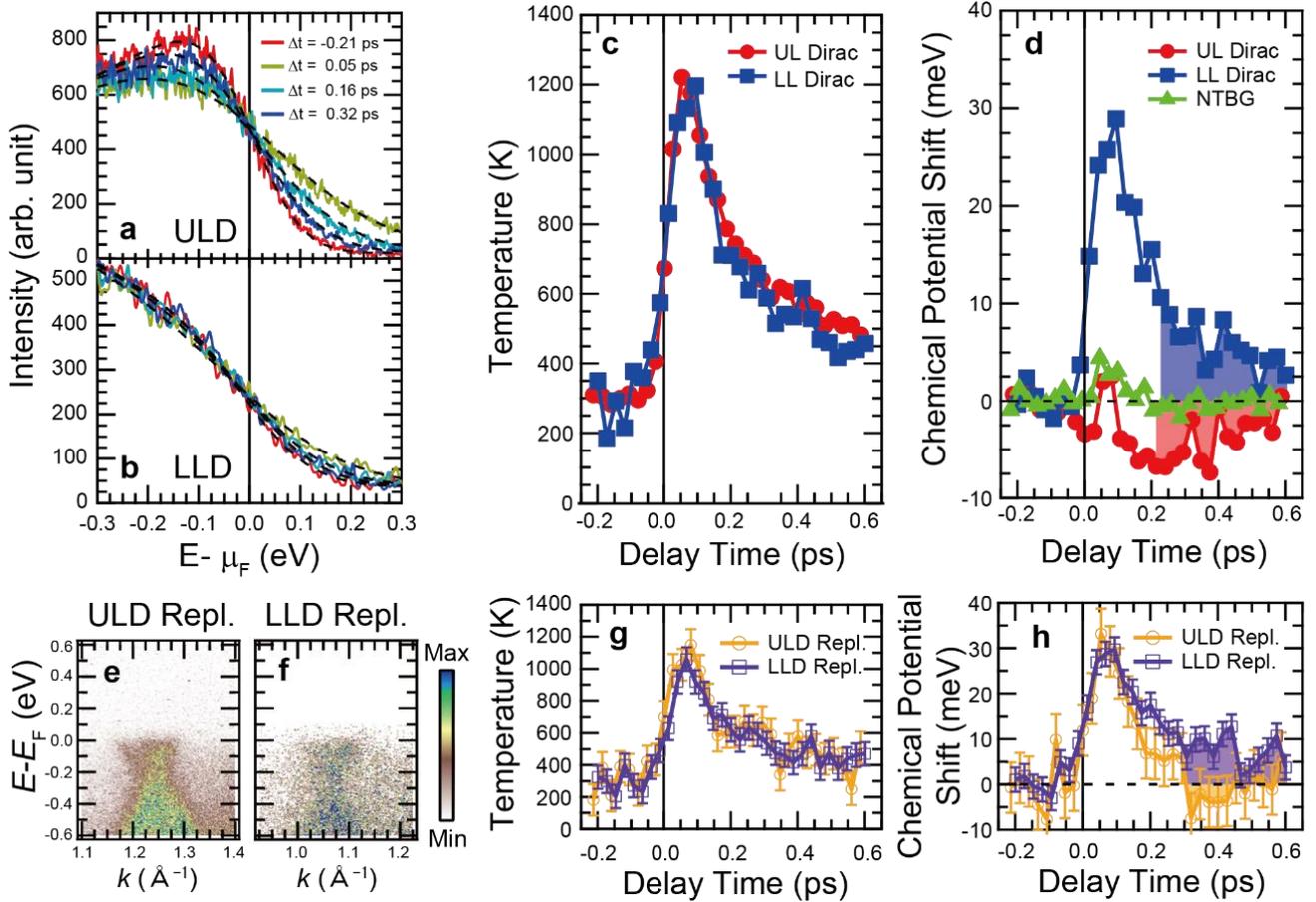

**Figure 3 | Time-dependence of electronic temperature and chemical-potential shift. a, b,** Energy-distribution curves for LLD and ULD bands as a function of pump-probe delays $\Delta t$, respectively. Dashed lines are fits to Fermi-Dirac distribution curves. **c,** Electronic temperature as a function of pump-probe delay time for LLD and ULD bands. **d,** Chemical-potential shift as a function of pump-probe delay time for LLD and ULD bands. For comparison, the result for NTBLG is also shown. **e, f,** Equilibrium angle-resolved photoemission spectroscopy map for replicas of the ULD and LLD. **g, h,** Electronic temperatures and chemical-potential shift as a function of pump-probe delay time for the

replicas of the ULD and LLD bands. Errors are estimated from the standard deviation in the data before the arrival of the pump pulse.

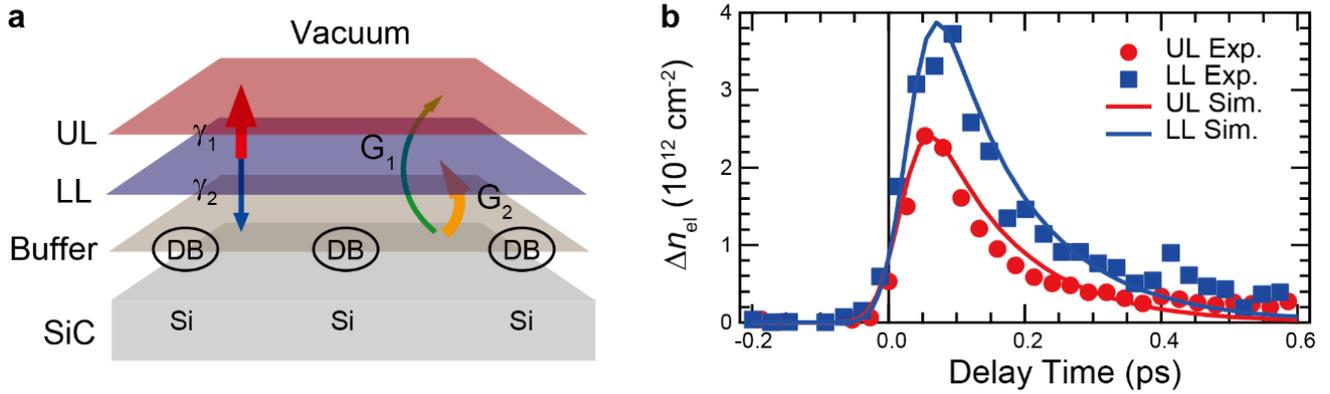

**Figure 4 | Carrier transport among layers in QCTBG**

**a,** Schematic illustration of the spatial relationship among the upper layer (UL), lower layer (LL), buffer layer, and SiC substrate in QCTBG. Schematic illustration of carrier transport among layers in QCTBG. $\gamma_1$ and $\gamma_2$ are rate constants of carrier transfer between the UL and LL, and the LL and SiC substrate. $G_1$ and $G_2$ are coefficients of pump-induced net density flux to the UL and LL from SiC substrate, respectively. Thicker lines indicate the larger values of $\gamma_1$ than $\gamma_2$ and $G_2$ than $G_1$, respectively. **b,** The change of electron densities, $\Delta n_{\mathrm{el}}$, as a function of pump-probe delay time for the UL and LL. Experimental results are shown as symbols whereas calculation results by solving rate equations are shown as lines.

# Supplementary Information

# Ultrafast unbalanced electron distributions in quasicrystalline 30° twisted bilayer graphene

T. Suzuki[1], T. Iimori[1], S. J. Ahn[2], Y. Zhao[1], M. Watanabe[1], J. Xu[1], M. Fujisawa[1], T. Kanai[1], N. Ishii[1], J. Itatani[1], K. Suwa[3], H. Fukidome[3], S. Tanaka[4], J. R. Ahn[2,5], K. Okazaki[1], S. Shin[1], F. Komori[1,*], and I. Matsuda[1,*]

[1]*Institute for Solid State Physics, University of Tokyo, Kashiwa, Chiba 277-8581, Japan*

[2]*Department of Physics and SAINT, Sungkyunkwan University, Suwon, Gyeonggi-do 16419, Republic of Korea*

[3]*Research Institute of Electronical Communication, Tohoku University, Sendai, Miyagi 980-8577, Japan*

[4]*Department of Applied Quantum Physics and Nuclear Engineering, Kyushu University, Fukuoka, Fukuoka, 819-0395, Japan*

[5]*Samsung-SKKU Graphene Centre, Sungkyunkwan University, Suwon, Gyeonggi-do 440-746, Republic of Korea*

# I. Umklapp scattering and replica bands

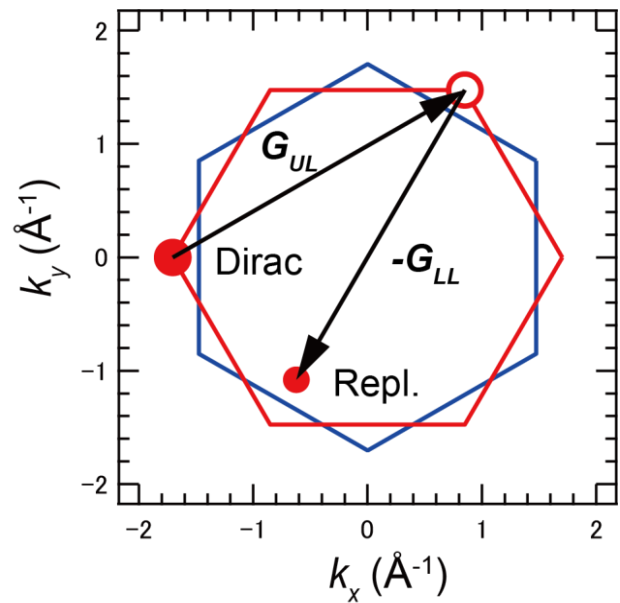

**Supplementary Figure 1 | Umklapp scattering and replica bands.** Red and blue hexagonal lines show the boundaries of the first Brillouin zone for the upper-layer (UL) and lower-layer (LL) graphene, respectively. The large (small) solid red circle represents the original (replica) band for the UL Dirac cone. $G_{UL}$ and $G_{LL}$ are reciprocal-lattice vectors of the crystal for the UL and LL, respectively.

## II. Crystal and electronic structures for quasicrystalline 30° twisted bilayer graphene and experimental setup

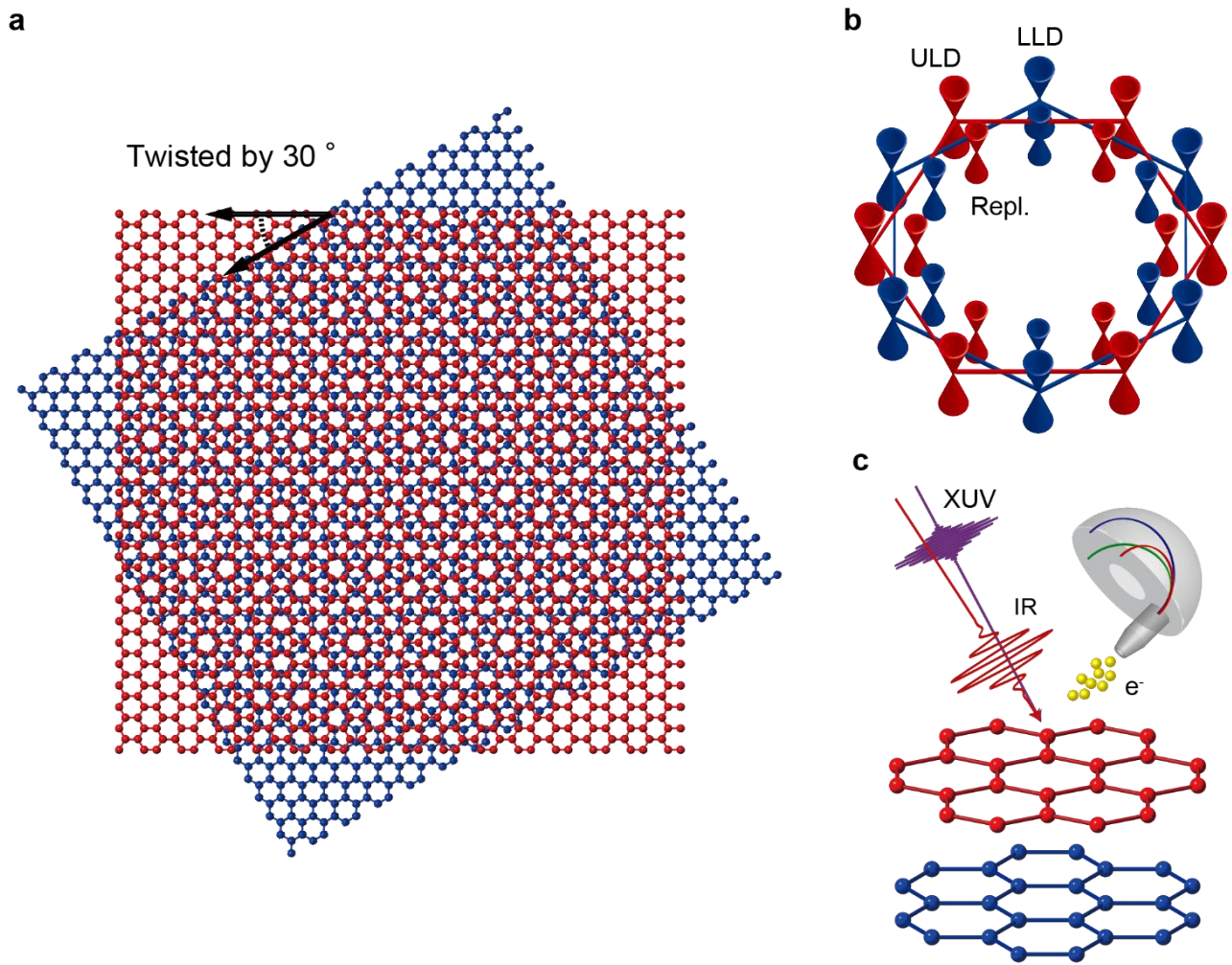

**Supplementary Figure 2 | Crystal and electronic structures of quasicrystalline 30° twisted bilayer graphene and experimental setup. a,** Crystal structure of quasicrystalline 30° twisted bilayer graphene (QCTBG). The upper- and lower-layer graphene sheets are twisted by 30° with respect to each other. **b,** Schematic drawing of electronic structures of QCTBG in the momentum space. Outer red and blue Dirac cones represent the upper-layer Dirac (ULD) and lower-layer Dirac (LLD) bands, respectively, whereas the inner red and blue Dirac cones corresponds to the replica bands of the ULD and LLD bands. **c,**

Schematic illustration of time- and angle-resolved photoemission spectroscopy (TARPES), as applied to QCTBG. The pump pulse is in the infrared whereas the probe pulse is in the extreme ultraviolet produced by high-harmonic generation. Photoelectrons are detected by using a hemisphere analyser.

## III. Model and calculation for nontwisted bilayer graphene

We describe in this section the model and calculation of the band structure for nontwisted bilayer graphene (NTBG), which depends strongly on the stacking order. Figures S3a and S3b show the crystal and band structure for AA- and AB-stacked NTBG. Two Dirac cones remain linearly dispersive and cross each other in AA-stacked NTBG, and the band becomes massive at the Dirac points in AB-stacked NTBG. Furthermore, depending on the doping, the band structure of AB-stacked NTBG becomes either gapped or gapless. For a quantitative description of these bands, we use the tight-binding model following Ref. [S1] to calculate the band structure. For AA-stacked NTBG, the Hamiltonian is

$$H_{AA} = \begin{pmatrix} 0 & \pi_k & t_{Int} & 0 \\ \pi_k^* & 0 & 0 & t_{Int} \\ t_{Int} & 0 & 0 & \pi_k \\ 0 & t_{Int} & \pi_k^* & 0 \end{pmatrix}, \quad (S1)$$

where $\pi_k = \frac{\sqrt{3}}{2} a t_{NN}(k_x + ik_y)$, $a$ is the lattice constant, $t_{NN}$ is the nearest-neighbour hopping parameter, and $t_{Int}$ is the interlayer hopping parameter. Figure S3C shows the calculated band structure, and Table S1 lists the values of each parameter used in this work [S2]. For AB-stacked graphene, the Hamiltonian is

$$H_{AB} = \begin{pmatrix} E_u & \pi_k & t_{Int} & 0 \\ \pi_k^* & E_u & 0 & 0 \\ t_{Int} & 0 & E_l - \frac{\Delta}{2} & \pi_k \\ 0 & 0 & \pi_k^* & E_l + \frac{\Delta}{2} \end{pmatrix}, \quad (S2)$$

where $E_u$ and $E_l$ are the energy for the UL and LL graphene sheets. Furthermore, the LL graphene experiences a sublattice asymmetry $\Delta$ due to interaction with the substrate. Figures S3d–S3f show the calculated results with $E_u = E_l = \Delta = 0$; $E_u = 0.5$ eV, $E_l = -0.5$ eV, $\Delta = 0$; and $E_u = 0.5$ eV, $E_l = -0.5$ eV, $\Delta = 0.2$ eV, respetively. In Fig. 2c, we use the results $E_u = 0.5$ eV, $E_l = -0.5$ eV, $\Delta = 0.2$ eV (Fig. S3f).

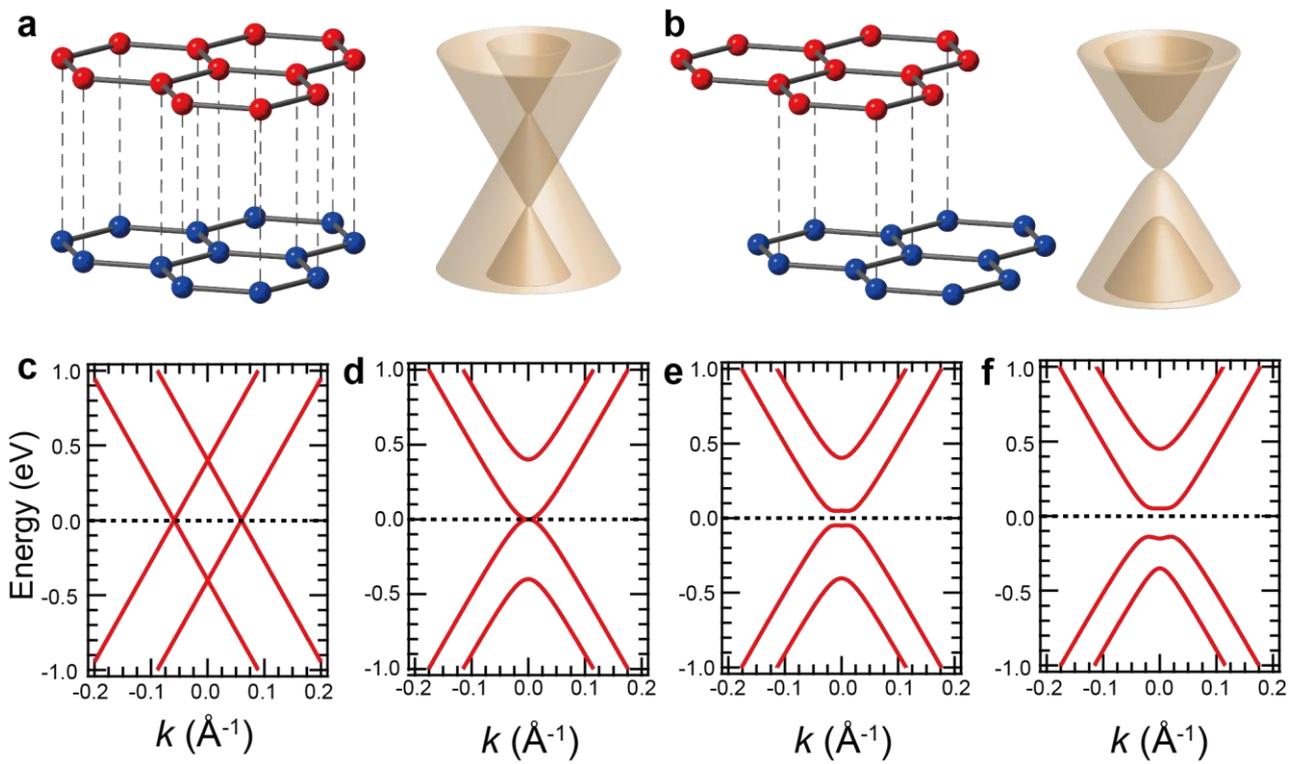

**Supplementary Figure 3 | Crystal and band structure for nontwisted bilayer graphene.**
**a,b,** Crystal and band structure for AA- and AB-staked bilayer graphene. **c–f,** Calculated band structure for (a) AA-staked and (b) AB-stacked bilayer graphene with (c), (d) $E_u = E_l = \Delta = 0$; (e) $E_u = 0.5$ eV, $E_l = -0.5$ eV, $\Delta = 0$; and (f) $E_u = 0.5$ eV, $E_l = -0.5$ eV, $\Delta = 0.2$ eV.

| Parameter | Value |
|---|---|
| $a$ | 2.46 Å |
| $t_{NN}$ | 3.16 eV |
| $t_{Int}$ | 0.4 eV |

Supplementary Table 1 | Parameters for NTBG used in calculations [S2].

IV. **Evaluation of time constant of time-dependent electronic temperature for each band in QCTBG.**

We fit the data to single-exponential functions convoluted with a 70-fs-wide Gaussian to obtain a full width at half maximum that corresponds to the time resolution of the experiment. The fits are shown as solid lines in Figs. S4a and S4b. The extracted time constants for each band in QCTBG are listed in Table S2.

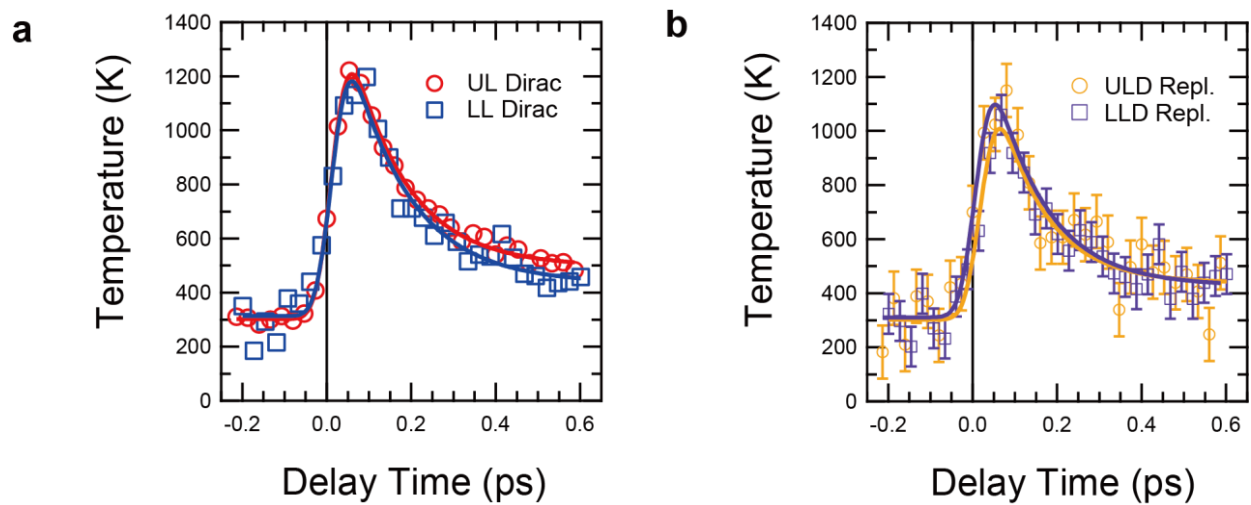

**Supplementary Figure 4 | Temporal evolution of electronic temperature for each band in QCTBG. a,** Time-dependent electronic temperature for the upper-layer (UL) and lower-layer (LL) Dirac bands in QCTBG. Data and fits are represented by symbols and lines, respectively. **b,** Results for the UL Dirac (ULD) and LL Dirac (LLD) replica bands.

| Band | Time constant |
|---|---|
| UL Dirac | 133±6 fs |
| LL Dirac | 141±20 fs |
| ULD Replica | 124±35 fs |
| LLD Replica | 114±22 fs |

Supplementary Table 2 | Time constant for time-dependent electronic temperature for each band in QCTBG.

## V. Rate equations for electron transport

A set of rate equations for electron transport is

$$\frac{dn_{el}^{SL}}{dt} = -\frac{n_{el}^{SL}}{\tau}, \qquad (S3)$$

$$\frac{dn_{el}^{UL}}{dt} = -\frac{n_{el}^{UL}}{\tau_{UL}} + D_1\left(n_{el}^{LL} - n_{el}^{UL}\right) + G_1 \exp\left(-\frac{t^2}{T_p^2}\right), \qquad (S4)$$

$$\frac{dn_{el}^{LL}}{dt} = -\frac{n_{el}^{LL}}{\tau_{LL}} - D_1\left(n_{el}^{LL} - n_e^{UL}\right) - D_2\left(n_{el}^{LL} - n_{el}^{Sub}\right) + G_2 \exp\left(-\frac{t^2}{T_p^2}\right), \qquad (S5)$$

$$n_{el}^{UL} + n_{el}^{LL} + n_{el}^{Sub} = const., \qquad (S6)$$

where $n_{el}^{SL}$, $n_{el}^{UL}$, $n_{el}^{LL}$ are electron densities in Dirac cones for a single-layer graphene, the upper-layer (UL), and lower-layer (LL) in a QCTBG, respectively. $\tau$ is a lifetime for a Dirac fermion in a single-layer graphene. $D_1$ and $D_2$ are diffusive coefficients between the UL and LL, and the LL and SiC substrate. Here, the SiC substrate includes the buffer layer and the Si dangling bonds (DB) at the interface between the SiC crystal and the bilayer graphene. $n_{el}^{Sub}$ is the electron density in the Si DB state at the graphene/SiC interface. $G_1$ and $G_2$ are coefficients of pump-induced net density flux to the UL and LL, respectively, where $T_p$ corresponds to the time resolution of our experiment (70 fs in the full width at half maximum). We also assume that the total electron density among the UL, LL, and SiC substrate is conserved, as represented in Eq. (S6).

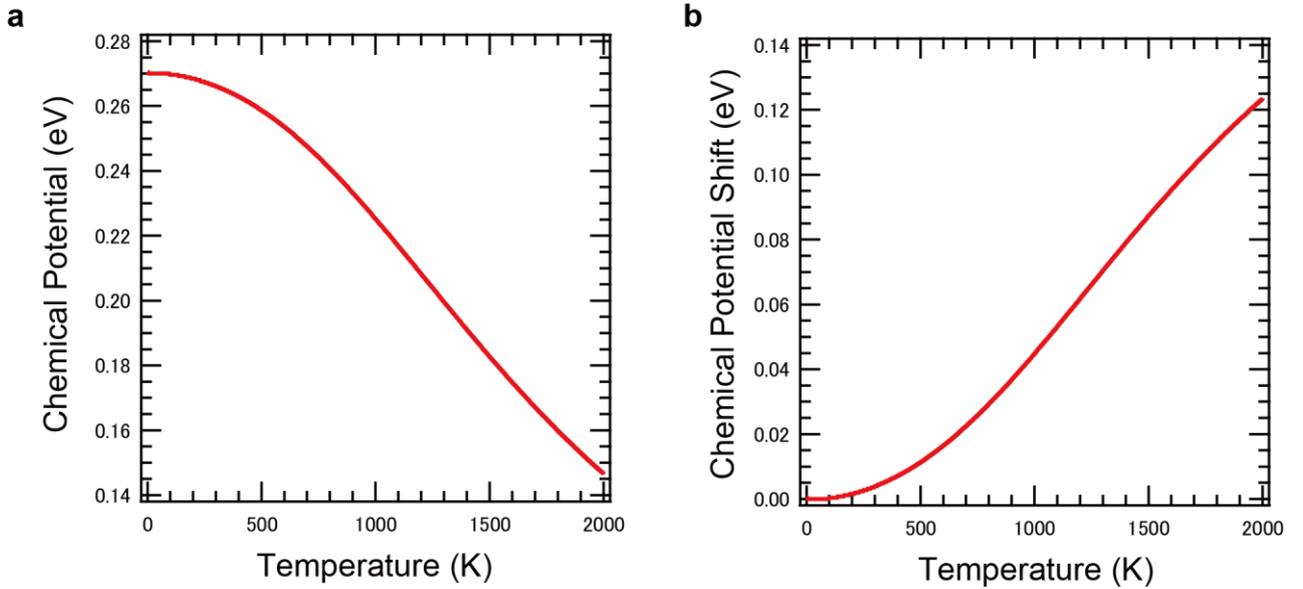

**Supplementary Figure 5 | The relation between chemical potential a,** Relation between temperature and chemical potential and **b,** chemical potential shift.

Before solving a set of rate equations, we will give the relationship between temperature, $T$, and chemical potential, $\mu$. Assuming the carrier conservation, $T$ and $\mu$ should suffice

$$\int_{-\infty}^{+\infty} \rho(E) f_{FD}(E, T, \mu) dE = const., \qquad (S7)$$

where $\rho$ and $f_{FD}$ are the density of states (DOS) for a graphene and the Fermi-Dirac distribution function, respectively. Figure S5 show the result of the relationship between $T$ and $\mu$.

We now calculate carrier transfer dynamics by solving rate equations. For Eq. (S3), we evaluate carrier dynamics in a single-layer graphene by hot carrier dynamics. We define hot carrier density, $n_{\text{Hot}}(T)$, as

$$n_{\text{Hot}}(T) = \int_{\mu(T)}^{+\infty} \rho(E) f_{FD}(E, T, \mu(T)) dE, \tag{S8}$$

where $\mu(T)$ is solely given by electron temperature via Eq. (S7). Figure S6a shows the spectral density as a function of energy. Blue-dashed and red-solid lines represent spectral densities for $T = 0$ and 1200 K, respectively. $n_{\text{Hot}}(T = 1200\text{K})$ corresponds to the hatched area shown in Fig. S6a, corresponding to Eq. (S8). Figure S6b shows the relationship between hot carrier density and temperature.

We convert the experimental results for electron temperature dynamics shown in Fig.3c to hot carrier densities, $n_{\text{Hot}}$, for the UL and LL as shown in Fig. S6c. Here, differences form pre-arrival of pump, $\Delta n_{\text{Hot}}(t) = n_{\text{Hot}}(t) - n_{\text{Hot}}(t < 0)$, are shown. Red and blue symbols correspond to time-dependent hot charrier densities for the UL and LL, respectively. By single exponential fits shown as black-dashed lines, hot carrier lifetime, $\tau$, are evaluated as 0.156, and 0.140 ps for the UL and LL, respectively, denoted as $\tau_{UL}$ and $\tau_{LL}$ in Eq. (S4) and (S5).

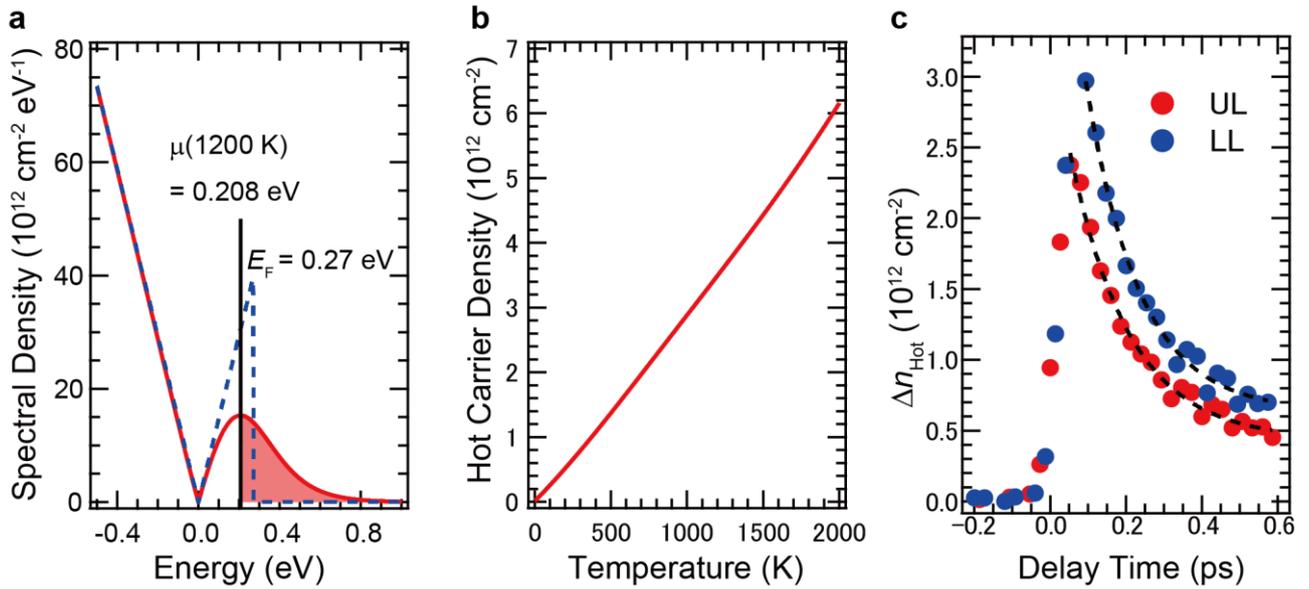

**Supplementary Figure 6 | Dynamics of hot carriers. a,** Spectral density as a function of energy. Blue-dashed and red-solid lines are spectral densities for $T = 0$ and 1200 K, respectively. Hot carrier density for $T = 1200$ K corresponds to the hatched area. **b,** Relationship between temperature and hot carrier density. **c,** Time-dependent hot carrier densities for the upper-layer (UL) and lower-layer (LL), shown as red and blue symbols, respectively. Black-dashed lines are single exponential fits.

Next, we proceed to solve Eqs. (S4) and (S5). Depending on parameters of $D_1, D_2, G_1,$ and $G_2$, dynamical behaviour of electron density $n_{el}(t)$ significantly changes. Figure S7 shows simulation results of the rate equations for time-dependent electron densities for the UL and LL Dirac shown as differences from pre-arrival of pump, $\Delta n_{el}(t) = n_{el}(t) - n_{el}(t < 0)$. Figures S7a and S7b show the results with varying $G_1$ while Figs S7c and S7d show the results with varying $D_1$. From these results one can see that $G_1$ and $G_2$ are mostly responsible for the initial dynamics, which raises electron densities, while $D_1$ and $D_2$ determine the relaxation dynamics.

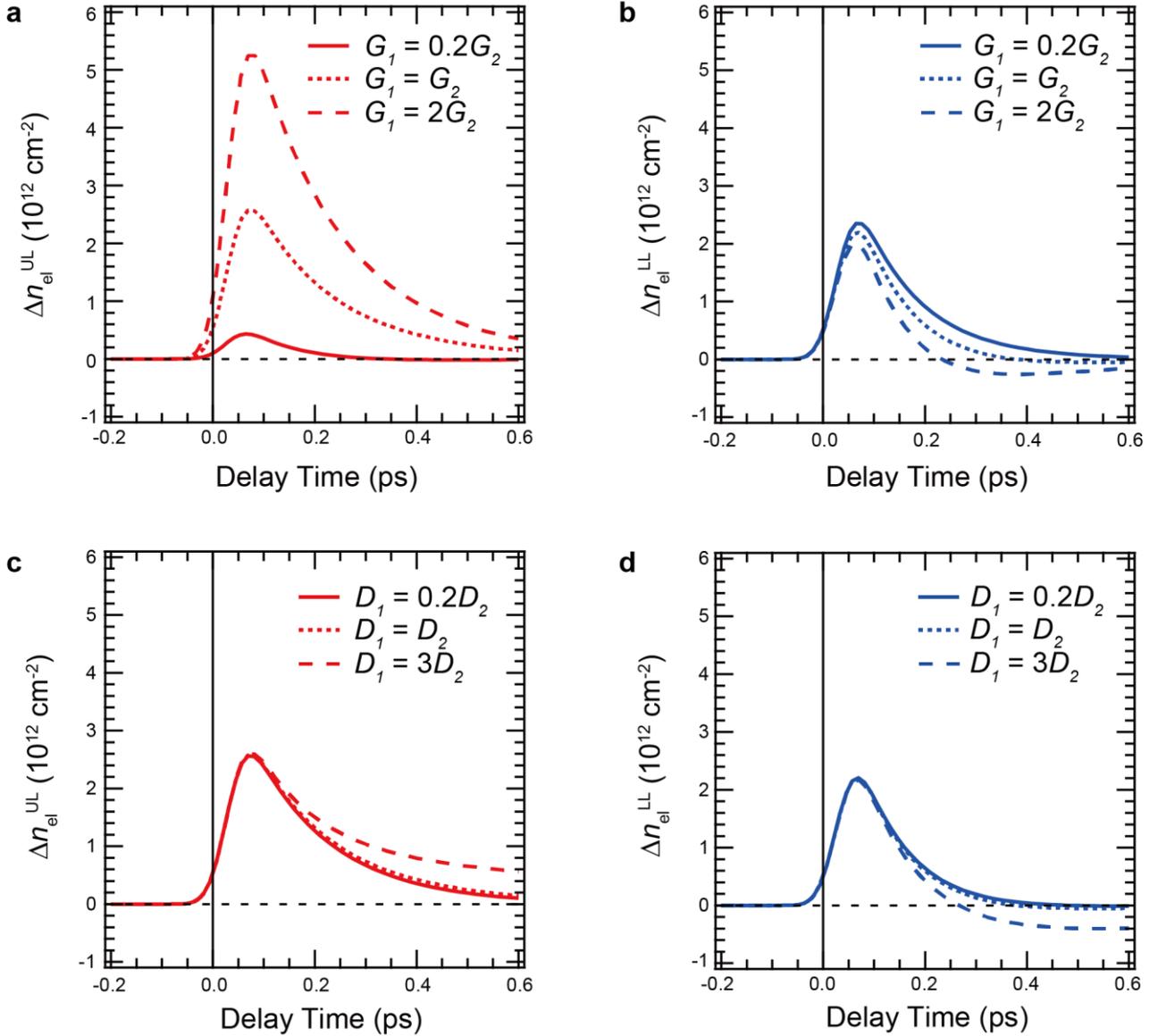

**Supplementary Figure 7 | Simulation results for rate equations. a, b,** Time-dependent electron densities for the UL and LL Dirac bands with varying value of $G_1$ while $G_2 = 5 \times 10^{13}$ cm$^{-3}$ps$^{-1}$, $D_1 = D_2 = 1.0$ ps$^{-1}$ are fixed. **c, d,** Time-dependent electron densities for the UL and LL Dirac bands with varying value of $D_1$ while $G_1 = G_2 = 5 \times 10^{13}$ cm$^{-3}$ps$^{-1}$, $D_2 = 1.0$ ps$^{-1}$ are fixed.

In order to obtain the experimentally-determined time-dependent electron densities $n_{el}(t)$ from the time-dependent electronic temperature $T(t)$ and chemical potentials $\mu(t)$, we use the relationship given by

$$n_{el}(t) = \int_0^{+\infty} \rho(E) f_{FD}(E, T(t), \mu(t)) dE. \quad (S9)$$

Figure S8 show time-dependent electronic temperature, chemical potential shifts, and electron densities. Electron densities are shown as differences from pre-arrival of pump, $\Delta n_{el}(t) = n_{el}(t) - n_{el}(t<0)$. The red- and blue-solid lines in Fig. S8c correspond to the calculation results of rate equations for the UL and LL bands, respectively. The used parameters of $D_1, D_2, G_1$, and $G_2$ are shown in Table S3.

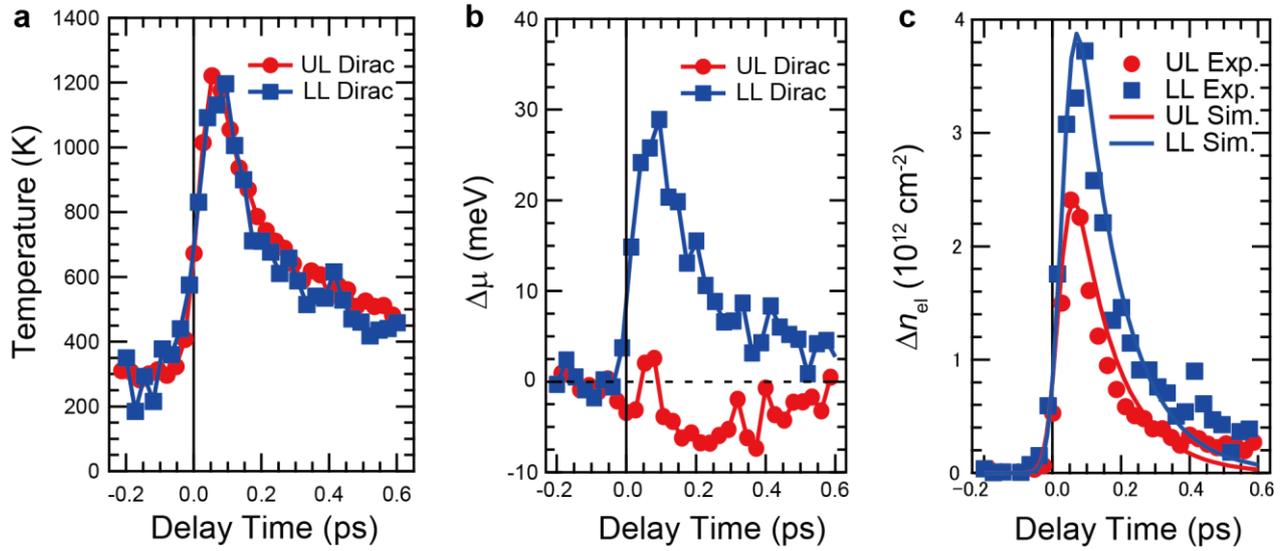

**Supplementary Figure 8 | Electron density dynamics and simulation results for rate equations. a, b,** Time-dependent electronic temperature and chemical potential shifts for the UL and LL Dirac bands. **c,** Time-dependent electron densities and simulation results for the UL and LL Dirac bands. Data and simulations are represented by symbols and lines, respectively.

| Parameter | Value |
|---|---|
| $D_1$ | 1.5 ps$^{-1}$ |
| $D_2$ | 0.5 ps$^{-1}$ |
| $G_1$ | $5 \times 10^{13}$ cm$^{-3}$ps$^{-1}$ |
| $G_2$ | $8 \times 10^{13}$ cm$^{-3}$ps$^{-1}$ |

Supplementary Table 3 | Parameters used in calculations of rate equations.